# Properties of Relationships among objects in Object-Oriented Software Design

*Zeynab Rashidi*

Master Student in Department of Mathematics and Computer Science, AmirKabir University of Technology, Tehran, Iran, Email: zeynabrashidi@aut.ac.ir

**Abstract**: One of the modern paradigms to develop a system is object oriented analysis and design. In this paradigm, there are several objects and each object plays some specific roles. After identifying objects, the various relationships among objects must be identified. This paper makes a literature review over relationships among objects. Mainly, the relationships are three basic types, including generalization/specialization, aggregation and association.This paper presents five taxonomies for properties of the relationships. The first taxonomy is based on temporal view. The second taxonomy is based on structure and the third one relies on behavioral. The fourth taxonomy is specified on mathematical view and fifth one related to the interface. Additionally, the properties of the relationships are evaluated in a case study and several recommendations are proposed.

## 1-Introduction

The modern paradigm for developing software is Object-Oriented (OO). In this paradigm, we describe our world using the object categories (classes) or object types (pure abstract class or Java interface) (see[12],[13]and [26]). Each class/object plays a specific role in the software. These roles are programmed in Object-Oriented languages such as C++ and Java.Severalattributes (data variables) and services (operations/functions/methods) are assigned to these classes. Then, we model the behavior of the world as a sequence of messages that are sent between various objects. In OO models, a number of relationships (inheritance, association, and aggregation- see [22],[3], [20], [23]and [26]) are identified between the classes/objects. Moreover, there are many popular design modeling processes and guidelines such as GRASP [28] and ICONIX [27] for assigning responsibility to classes and objects in object-oriented design.

In recent years, few researchers focus on object oriented software engineering. Fokaefs et al. (2012) describe a method and a tool designed to fulfill exactly the extract class refactoring [11]. The method involves three steps: (a) recognition of extract class opportunities, (b) ranking of the opportunities in terms of improvement to anticipate which ones to be considered to the system design, and (c) fully automated application of the refactoring chosen by the developer. Bavota et al. (2014) propose an approach for automating the extract class refactoring [1]. This approach analyzes structural and semantic relationships between the methods in each class to identify chains of strongly related methods. The identified method chains are used to define new classes with higher cohesion than the original class, while preserving the overall coupling between the new classes and the classes interacting with the original class.

The first step for building an OO model is to find out the objects. In this step, we are not really finding objects. In fact, we are actually finding categories and types (analysis concepts) that will be implemented using classes and pure abstract classes. The results of problem analysis is a model that: (a) organizes the data into objects and classes, and gives the data a structure via relationships of inheritance, aggregation, and association; (b) specifies local functional behaviors and defines their external interfaces; (c) captures control or global behavior; and (d) captures constraints (limits and rules).

In the real world, no object could not be independent of all other objects, similar to an island. Objects typically depend on other objects for services and possibly for error handling, constant data, and exception handling. Relationships capture the interdependencies between objects and provide the means by which objects know about each other. In object orientation, every service request (function call) must be sent to a specific object while in the procedural languages a function can be called directly.For example, in order for object A to send a message to object B, object A must have a handle to object B (in C++, a reference or pointer).Accessing another object's services can be performed in the following ways(See [7], [9], [10], [11], [14] and [29]):

- The calling object, which has a handle, passes the handle of the other object as one of the arguments of the function (message) signature.
- The called object has a relationship(aggregation or link) to the other object.
- The needed service belong to an 'ancestor' class. Ancestor means a super-class.
- The access of static class function, which may be considered a managed global function.

The main motivation of this paper is to survey the relationships among objects and makes five taxonomies for their properties. The structure of remaining sections is as follows. In Section 2, the literature review and main relationships among objects are described. In Section 3, the taxonomies are specified. In Section 4, practical experience and guidelinesare presented. Finally, Section 5 is considered to summary and future works.

## 2-Literature Review

In the literature ( [2], [3], [4], [5], [6], [8], [16], [26], [23] and [20]), we found three basic relationships among classes/objects: generalization/specialization (inheritance),aggregationand association.These are certainly not new concepts and most professionals work with them every day in modeling.

- **Generalization/Specialization**:We all learned generalization/specialization when studying taxonomies inbiology class. This is a relationship between classes rather than objects. Generalization/Speciation 'Is A Type' relationship between classes. For example, consider two objects: 'Person' and 'Student'. Student 'is-a' Person. Thus, the attributes of a person is also attributes of student. In this relationship, attributes, relationships, services, and methods are inherited from the generalization (super-class) by the specialization (subclass).
- **Aggregation**:This is a relationship in which one object is formed from other objects; e.g. Car and engine.Aggregation captures the whole-parts relationship between objects. In contrast to generalization/specialization, there is no inheritance between objects participating in an aggregation. The main advantages of aggregations are that they reduce complexity by allowing software engineers to treat many objects as one object.
- **Association**: This is a relationship by which an object knows about another one. An excellent example of an association (link) is marriage. Moreover, links in the form of associations have been widely usedfor yearsin the database modeling community.

These relationships and their identifications are described in the following subsections.

## 2-1-Generalization/Specialization

To identifying generalization/specialization relationship, software engineers must perform the '**IS_A**'test between pairs of objects after identifying objects. In fact, software engineers ask the questions: (a) Is Object A an Object B? ; (b) Is Object B an Object A? Note that we are really asking if an object of type A is an object of type B. Allowed answers of those questions are: (a) 'always'; (b) 'sometimes' and (c) 'never'. Based on the answers, software engineers make some interpretations according to the information given in Table-1. More details on the matter along with some examples are given in [16].

Table-1: The interpretation of the results in IS_A Test

|  | Questions | | Interpretation |
|---|---|---|---|
|  | Is B an A? | Is Aa B? |  |
| Answer | Always | Always | Synonymous |
|  | Sometimes | Always | B is a generalization of A |
|  | Always | Sometimes | A is a generalization of B |

## 2-2- Aggregation

Unfortunately, most software engineers have difficulties applying this relationship properly in practice because the object-oriented paradigm has not defined the aggregation mechanism very well. The latest literature on this topic argues that this is due to the fact that aggregation, itself, is an 'ancestor' concept. It is our belief that software engineers need to use the 'descendent' concepts (more specialization) to be able to use this mechanism effectively. These descendent concepts, or different kinds of aggregation, will capture additional properties that will help software engineers to manage complexity effectively.

From a theoretical perspective, linguists, logicians, and psychologists have studied the nature of relationships. One of relationships that has been studied reasonable well is the relationship between the parts of things and the wholes that they make up. In a joint paper, Morton Wins ton, Roger Chaffin, and Douglas Herrmann discussed this whole-parts relationship[25]. They described several kinds of aggregation. The paper identified six types of aggregation; Lee and Tepfenhart (2005) added a seventh[16]; we added an eighth:(a)Assembly-Parts;(b) Component-Integral Composition;(c) Material-object Composition;(d) Portion-Object Composition;(e) Place-Area Composition;(f) Collection-Members composition;(g) Container-Content(Member-Bunch Composition);(h) Member-partnership composition and (i) Compound-Elements Composition.

- **Assembly-Parts (Component-Integral) Composition**: In this aggregation, the whole is comprised of the components that maintain their identity even when they are part of the whole. The parts have a specific functional or structural role with respect to each other. To identify this aggregation, software engineers must look for some keywords like 'is part of' and 'is assembled from'. For example, a keyboard is part of a computer and chairs are parts of the office. Note that in this relationship the assembly does not exist without parts and the components may not be haphazardly (incidentally) arranged, but must bear a particular relationship, either structurally or functionally. Moreover, the whole exhibits a patterned structure or organization. In practice, the whole may be: (a)Tangible like car, toothbrush and printer; (b) Abstract like physics, mathematics and jokes; (c) Organizational like NATO and United Nation; (d) Temporal such as musical performance and film showing.
- **Material-Object Composition**: In this type of aggregation, the parts lose their identity when they are used to make the whole. This defines an invariant configuration of parts within the whole because no part may be removed from the whole. To identify this relationship, software engineers must look for key words like 'is partly' and 'is made from'. For example, suppose bread is made from flour, a table is made from wood and a car is made of materials such as iron, plastic and glass.
- **Portion-Object Composition**: This aggregation defines a homogenous configuration of parts in the whole. Usually, portions of the objects can be divided using standards

- measures such as inches, liters, hours and so on. The portion-object composition supports the arithmetic operations +, -, ×, /. To identify this kind of relationship, software engineers must look for some keywords like 'portion of', 'slice', 'helping of', 'segment of', 'lump of', and such similar phrases. For example, a second is part of a day and a meter is part of kilometer.
- **Place-Area Composition**: This aggregation defines a homogenous and invariant configuration of parts in a whole. It is commonly used to identify links between places and particular locations within them. When looking for this aggregation, look for preliminary Portion-object composition and then ask if this relationship is invariant. For examples, Colchester is part of UKand a room is part of a hotel.
- **Collection-Members Composition**: This aggregation is a specialized version of the Place-Area Composition. In addition to being homogenous and invariantconfiguration of parts within a whole, there is an implied order to its members. When looking for this aggregation, look for place-area aggregation and then check if there is an implied order. For example,suppose Collection-members Composition Airline reservation with its various flight segments and Monthly timesheet-daily timesheets.
- **Container-Content (Member-Bunch)Composition**: This aggregation defines a collection of parts as a whole. The only constraint, here, is that there is a spatial, temporal or social connection for determining when a member is part of the collection. This aggregation tends to be a catchall (contents without classification) for aggregation-type relationships.For example, suppose a box with contents of the box and a bag with its contents of bag.
- **Member-Partnership Composition:**In this aggregation, the parts bear neither a functional nor a structural relationship to each other or to the whole. The contents are neither homogenous nor invariant. For example, we can consider an Union and members and a Company and its employees. This is an invariant form of the container-content aggregation. Members in this relationship cannot be removed without destroying the aggregation.
- **Compound-Elements Composition:** In this aggregation, the parts bear neither a functional nor a structural relationship to each other or to the whole. The contents are homogenous and variant. The components arehaphazardly (incidentally) arranged in the whole. For example, we can consider a Party and People in a society.

Note that an object can be viewed as more than one aggregation. For example, we can consider Bread as aggregate of slices (Portion-Object) and Bread as made of flour, egg(Material-Object).

**2-3-Association**
An association is a relationship that allows an object to know about another one. This relationship is considered to be bi-directional as link through which one object traverses in either direction. An association can have attributes and services. The best source for initial identification and specifying associations and aggregations is the requirements documents. Links, like services are often seen as verbs. For example, 'which it gets from', 'keep track of', 'changes with', and 'depends upon'. The sequence diagrams and behavior specification documents also help to find the links.

When software engineers are distinguishing between association and aggregation, several points must be considered: (a) An aggregation may not connect an object to itself (e.g., supervise is between two instances); (b) Multiple connections between objects are possible (e.g. Employee doing several tasks). (c) Self associations are possible and common (e.g.

Sibling association on Student) and (d) Multiple association does not imply that the same two objects are related twice.

**3- Taxonomies**
One the major gaps and research needs is to have an overview and taxonomy on properties of relationships among classes/objects in Object-Oriented software development. According to Merriam-Webster[18], taxonomy is the study of the general principles of scientific classification, and is especially the orderly classification of items according to their presumed natural relationships. The major differences between properties of relationships among objects, in general, depend on the temporal, structure, behavioral and interface views, and in particular mathematical view. There are, therefore, five taxonomies to categorize properties of the relationships among objects in Object-Oriented development. These taxonomies are described in the following sub-sections.

**3-1-The First Taxonomy: Properties on Temporal View**
The first taxonomy for properties of the relationships among objects is concerned with varying aggregation dependency over time. Therefore, there are two properties of the relationship in this taxonomy:
- **Static**: In this property, components in a whole are fixed and cannot be changed over time. In the aggregations specified in Section 2-2, Assembly-Parts(Component-integral) Composition, Material-object Composition and Portion-object Compositionare in this taxonomy. For example, a telephone is assembled from its parts and Windows are parts of a house.
- **Dynamic**: In this property, components in a whole may vary over time. In the aggregations identified in Section 2-2, Material-Object Composition,Place-area composition, Collection-members composition, Container-content (Member-Bunch) composition and Member-Partnership composition are dynamic.

**3-2-The Second Taxonomy: Properties on Structure View**
The second taxonomy is based on the question of whether or not the relationships bear a particular functional or structure among classes/objects. In the generalization/specialization relationship, this taxonomy related to the following properties:
- **Attributes:** The descendent will have all of the attributes of the ancestor. For instance, suppose the Employee class that inherits from the Person Class in a general payment system; The Employee has the age attribute because it is a descendant class of Person.
- **Links:** The descendant will have all of the non-generalization links of the ancestor. For example, if we add a marriage link between two persons, 'Student' will also have a marriage link because it is a descendent of 'Person'.

In the aggregation relationship, we can categorize the properties of the relationshipsaccording to the combination of the following aspects:
- **Configuration**: In this aspect, we must determine whether or not the parts bear a particular functional or structure relationship.
- **Homogenous**:In this aspect, we determine whether or not the parts are from the same kind of thing in the whole.
- **Invariance**:In this aspect, the kind of the relationship is determined by the basic properties of whether or not the parts can be separated from the whole.

Table-2 shows the types of aggregation identified in Section 2-2, according to the propertieson the structure view.

Table-2: Different combination of properties in the Aggregation relationship

| Type of Aggregation | Configuration | | Homogenous | | Invariance | | Example |
|---|---|---|---|---|---|---|---|
| | Yes | No | Yes | No | Yes | No | |
| Assembly-Parts (Component-Integral) Composition | √ | | √ | | | √ | Windows are parts of a house |
| Material-Object Composition | √ | | √ | | √ | | A car is made of materials such as iron, plastic and glass |
| Portion-Object Composition | √ | | √ | | | √ | A second is part of a day |
| Place-Area Composition | √ | | √ | | √ | | A room is part of a hotel |
| Collection-Members Composition | | √ | | √ | √ | | Monthly timesheet and daily timesheets |
| Container-Content (Member-Bunch) Composition | | √ | | √ | | √ | A box and contents of the box |
| Member-Partnership | | √ | √ | | | √ | Union and members |
| Compound-Elements Composition | | √ | √ | | | √ | A party and several people |

### 3-3-The Third Taxonomy: Properties on Behavior View

The third taxonomy for properties of the relationshipsis based on how the behavior of classes/objects depending on others. In Generalization/Specializationrelationship, we have two types of properties:

- **Generalization without polymorphism (Good child):** All methods supplied by the ancestor for services are also used by the descendent to provide the corresponding services.
- **Generalization with polymorphism (Bad child):** Some methods provided by the ancestor for its services are used by the descendant. However, the descendant can supply its own customized methods that replace the appropriate methods.

### 3-4-The Fourth Taxonomy: Properties on Mathematical View

The fourth taxonomy for properties of relationships is based on mathematical view. In the generalization/specialization relationship, we have two following properties between classes:

- **Anti-symmetric**: If class A is a descendant of class B, then class B can not be a descendant of class A. e.g. 'Employee' is a person, but not all persons are employees.
- **Transitivity:** If class C is a descendant of class B and Class B is a descendant of Class A, then Class C is a descendant of Class A. e.g. if we add the fact that a 'Salesperson' is a 'Employee' then 'SalesPerson' is also a 'Person'. Furthermore, it also has the age attribute.

In the aggregation relationship, we have two following properties of the relationship between objects:

- **Anti-symmetry**: If an object A is a part of an object B, then the object B cannot be a part of the object A.
- **Transitivity**: If an object C is part of an object B and the object B is part of an object A, then C is part of A.

Note that the transitivity holds only for aggregations of the same kind. For example, we can consider: (a) Microwave is part of a kitchen (Component-integral) and (b) Kitchen is part of a house (Place-area), but Microwave is not part of a house.

### 3-5-The Fifth Taxonomy: Properties on Interface View

The fifth taxonomy for properties the relationshipsis related to providing service by an object for others.With this view, in the generalization/specializationrelationship the descendant must also provide all services provided by the ancestor. For example in a Personnel Management System, if the 'Person' object had a 'Get_Degree' service, then 'Student' will also have a 'Get_Degree' service because 'Student' is a descendent of 'Person'.

In the association relationship, a link can be binary (between two objects), ternary (among threeobjects), or higher. In practice, it is rare to find links with a semantic meaning that tietogether objects of three different object types (classes)[16]. A good example for binary association would be a link between 'Student' and 'Course'. By extending this relationship, we can have a ternary relationship among the 'Student', 'Software', and 'Course' objects. It captures the fact that students use various software tools for different courses.

## 4-Practical Experience and Guidelines

In order to evaluate the relationshipsand their properties in practice, we used a Control Command Police System (CCPS) for which a mini-requirement is briefly described in [23].We expanded this system and used in our studydue to its fertility for reusability in both application and system software.This police service system must respond as quickly as possible to reported incidents and its objectives are to ensure that incidents are logged and routed to the most appropriate police vehicle. The most important factors that must be considered which vehicle to choose to an incident include:

- **Type of incident**: some important and worsening events need immediate response. It is recommended that specified categories of response actions are assigned to a definite type of incident.
- **Location of available vehicles**: Generally, the best strategy is to send the closest vehicle to address the incident. Keep in mind that it is not possible to know the exact position of the vehicles and may need to send a message to the car to determine its current location.
- **Type of available vehicles**: some incident need vehicles need and some special incident such as traffic accidents may need ambulance and vehicles with specific equipment.
- **Location of incident**: In some areas, sending only one vehicle for response is enough. In other areas, may be a police vehicle to respond to the same type of accident is enough.
- **Other emergency services such as fireman and ambulance**: the system must automatically alert the needs to these services.
- **Reporting details**: The system should record details of each incident and make them available for any information required.

The Use Case Diagram and Activity Diagram of this system are depicted in Fig. 1, and Fig. 2, respectively. We implemented this system in Microsoft Foundation Classes (MFC) as application framework for MS Windows (see [17], [19], [21], [27] and [28] for guidelines of implementations). The Class Diagram of this system is depicted in Fig. 3. In this class diagram, there are many classes. The main classes, here, are 'Incident', 'Police Staff', 'Police Vehicle', 'Police Officer', 'Director', 'Route Manager', 'Incident Waiting List', 'Response'and 'GPS Receiver'.

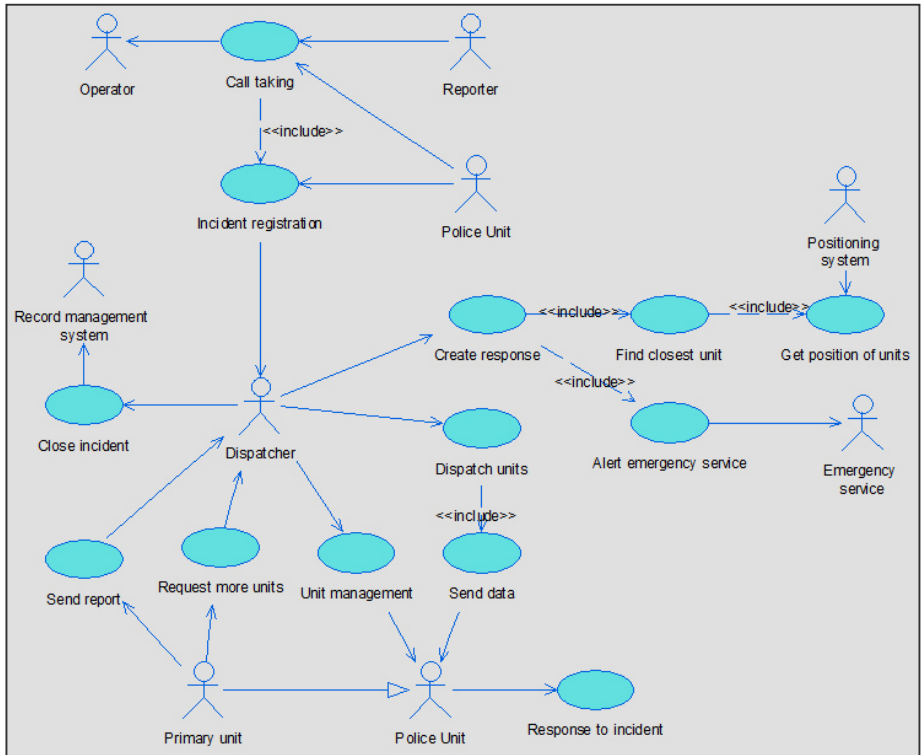

Fig. 1: The Use Case Diagram of the *Control Command Police System*

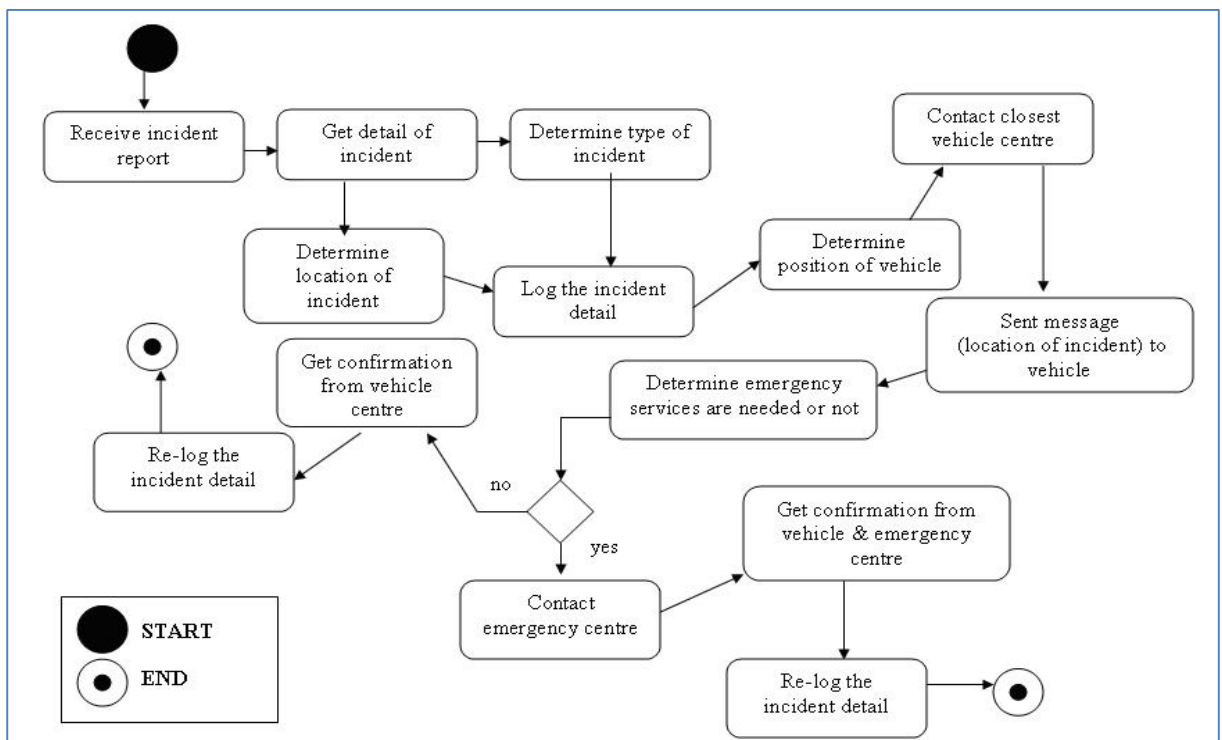

Fig. 2: The Activity Diagram of the *Control Command Police System*

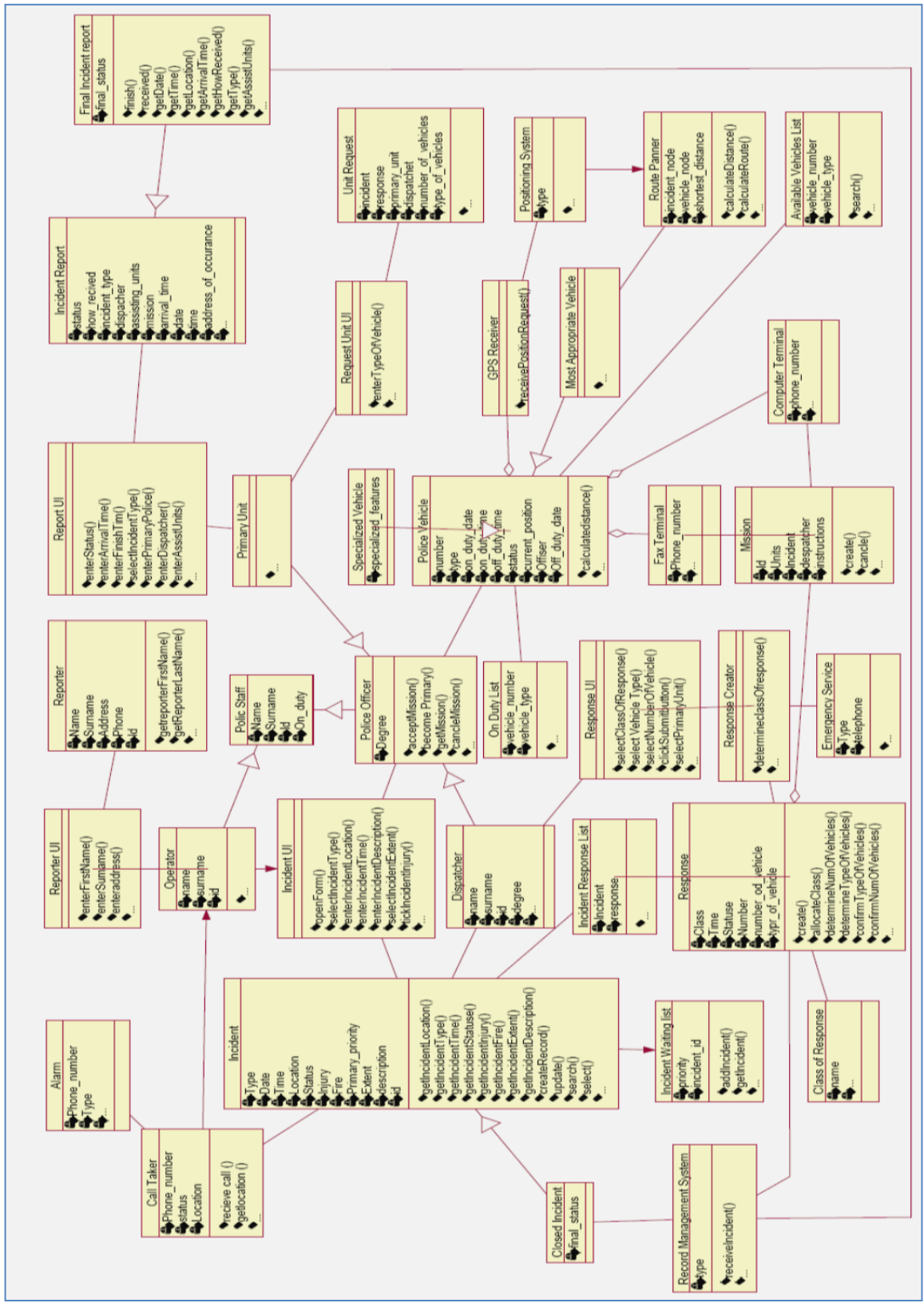

Fig. 3.The Class Diagram of the *Control Command Police System*

According to the experience, one of the most difficult tasks in building an object-oriented model is to determine whether a potential relationship is better captured as either an argument in the signature of the service (function), or as a link, aggregation, or generalization/ specialization. The following are the guidelines obtained from our experience:

- **Guidline-1**: A relationship must capture some concepts that applies to the problem domain or some sub-domain that is needed for implementation. In other words, there must be a semantic meaning to the relationship. A service (see the property on Interface view in Section 3-5) should only traverse the relationship when its usage is consistent with that semantic meaning. For example, consider the link between 'Specialized Vehicle' and 'Police Vehicle' (see Fig. 3). Today, with some security service, it is possible for 'Specialized Vehicle' to work for 'Police Vehicle'. It would be improper and poor modeling to use the link relationship to get to work domain services of the other vehicle. A second link (Security Service) needs to be established to capture this different semantic relationship.
- **Guidline-2**: When the relationship is 'permanent' (the static property in the first taxonomy in Section 3-1), software engineers must care around this term. If software engineers consider a scenario as a unit of time (e.g. across an incident in our experience), then permanent means that the relationship needs to be known across scenarios. Basically, if it has to be stored in memory for use by some other independent process like the management process between 'Dispatcher' and 'Police Office', then it is permanent.
- **Guidline-3**: In each aggregation, software engineers must make sure that all of the parts are in the same domain and provide the same functional or structural configuration to the whole. Apply transitivity and anti-symmetric properties tests (see the properties in Section 3-4) to check for consistency. Note that transitivity is possible only with aggregations of the same kind. It is very common for novices to mix parts of different kinds of aggregation in one aggregation. This will cause the transitivity test to fail. When this happens, software engineers probably need to look at the parts to see if there are different types of aggregates. For example, consider the Control Room that has the following parts: computer, monitors, printers, chairs, windows, floors, ceilings and walls. If we put all of these parts into one aggregation, we have mixed parts from two different semantic aggregations. The computer, monitors, printers are defining a functional configuration of the building; while the windows, floors (meaning the physical floor), ceilings, and walls are defining a structural configuration of the building. These parts must be captured in two different aggregations, as they have different semantics.
- **Guidline-4**: No aggregations connect two objects of the same kind to each other. This would violate the anti-symmetric property of the aggregation. For example in our experience, a 'Dispatcher' may not be an aggregate of 'Police officer'.
- **Guidline-5**: An association may connect two objects of the same kind. For example, the relation between the 'Reporter' and 'Reporter UI' in the Control Command Police System is valid (See Fig.3).
- **Guidline-6**: Aggregation is often confused with topological inclusion. In Topological inclusion, we have a relationship between a container, area, or temporal duration and that which is contained by it. Suppose in the Control Command Police system: (1) the 'Dispatcher' is in the control room, (2) the 'Incident' is in the evening, and (3) The 'Incident' is in Colchester and Essex. In each case, the container surrounds the subject. However, it is not part of the container in any meaningful semantic domain. For example, the 'Dispatcher' is not part of the control room, nor the 'Incident' is not a part of the evening. Furthermore, the 'Incident' is not part of Colchester or Essex.
- **Guidline-7**: The attributes of an object, sometimes, may be confused with aggregation. Attributes describe the object as a whole like a black box approach while aggregation describes the parts thatmake the whole similar to white box approach. In our experience of the Control Command Police system (see Fig. 3), the 'Route Planner' have attributes such as'Incident_Node' and 'Vehicle_Node'.

- **Guidline-8**:Attachment of one object to another object does not guarantee aggregation. Certainly'GPS Receiver' is attached to the 'Police Vehicle' and they are part of the system; however, 'Vehicle Radio' or 'Vehicle Stereo' are attached to the Vehicle, but they are not part of the Vehicle. Note that 'GPS Receiver' providefunctional support to the 'Police Vehicle', while 'Vehicle Radio' or 'Vehicle Stereo' do not supply any functional orstructural support in our case study.
- **Guidline-9**:Ownership may sometimes be confused with aggregation. Certainly a 'Police Vehicle' has a number,and 'GPS Receiver' are part of'Police Vehicle'. However, the fact that 'Dispatcher' has a vehicle does not implythat the 'Police Vehicle' is part of 'Dispatcher'. Thus, ownership must be captured by a link.
- **Guidline-10**:Multiple associationsamong objects are possible in which each association should be used to capture a distinctsemantic meaning. For example, the 'Alarm' and 'Call Taker' have multiple links in our experience (See Fig. 3).

## 5-Summary and Conclusion

This paper reviewed therelationships among objects in object-oriented software development and made five taxonomies for their properties.Mainly, the relationships are three basic types. This paper presents five taxonomies for properties of the generalization/specialization, association and aggregation relationships. The first taxonomy is based on temporal view and the second one is based on structure. The third taxonomy relies on behavioral view and the fourth one is specified on mathematical view. Finally, the fifth taxonomy related to the interfaces between objects. Moreover, in this paper the relationships are evaluated in a case study and then several recommendations are proposed.The main conclusion is that the relationships must capture some concepts that applies to the problem domain or some sub-domain. They are importantfor software engineers in implementation.